\documentclass{article}
\usepackage[letterpaper, margin=1in]{geometry}

\usepackage{booktabs}
\usepackage[utf8]{inputenc}
\usepackage[hyphens]{url}
\usepackage{hyperref}
\usepackage{graphicx}

\newcommand{\code}[1]{\texttt{#1}}

\usepackage[inline]{enumitem}

\usepackage[frozencache,cachedir=.]{minted}

\begin{document}

\title{Anomaly Detection in a Large-scale Cloud Platform}

\author{
Mohammad Saiful Islam$^*$, William Pourmajidi$^*$, Lei Zhang$^*$, \\ 
John Steinbacher$^\dag$, Tony Erwin$^\ddag$, and Andriy Miranskyy$^*$ \\
$^*$ Department of Computer Science, Ryerson University, Toronto, Canada \\
$^\dag$ Cloud Platform, IBM Canada Lab, Toronto, Canada \\
$^\ddag$ Watson and Cloud Platform, IBM, Austin, USA \\
\{mohammad.s.islam, william.pourmajidi, leizhang\}@ryerson.ca, \\
jstein@ca.ibm.com, aerwin@us.ibm.com, avm@ryerson.ca
}

\date{}

\maketitle

\begin{abstract}
Cloud computing is ubiquitous: more and more companies are moving the workloads into the Cloud. However, this rise in popularity  challenges  Cloud service providers, as they need to monitor the quality of their ever-growing offerings effectively. To address the challenge, we designed and implemented an automated monitoring system for the IBM Cloud Platform. This monitoring system utilizes deep learning neural networks to detect anomalies in near-real-time in multiple Platform components simultaneously. 

After running the system for a year, we observed that the proposed solution frees the DevOps team's time and human resources from manually monitoring thousands of Cloud components. Moreover, it increases customer satisfaction by reducing the risk of Cloud outages. 

In this paper, we share our solutions' architecture, implementation notes, and best practices that emerged while evolving the monitoring system. They can be leveraged by other researchers and practitioners to build anomaly detectors for complex systems.

\end{abstract}

\section{Introduction}
The way we have looked at things for decades has changed with digitalization. Our modern lifestyle depends on computers, electronic gadgets, the Internet and various Internet of Things (IoT) devices. In meeting the rapidly increasing technological demand, Cloud computing systems play an essential role. Cloud services are offered by remote servers and shared by customers on the Internet.  
As governments and companies move their facilities to Cloud systems, Cloud services are becoming increasingly popular. The market value of public Cloud services in 2019 was $\approx$~\$228 billion globally and is predicted to reach $\approx$~\$355 billion by 2022, with a 24\% year-over-year growth~\cite{GartnerF38:online}. Thus, Cloud systems are becoming critical to their users.

The backbone of a Cloud system and services it offers is a Cloud platform developed and maintained by the vendor (e.g., AWS platform is developed by Amazon, IBM Cloud --- by IBM, and Google Cloud --- by Google). The platform provides and controls a list of Cloud services, governs these services' deployment,  performs identity and access management, and handles business processes (e.g., client billing).  

A Cloud system consists of a huge number of hardware and software components. 
Appropriate health monitoring of these components and accurate detection of any abnormality play a vital role in ensuring reliable and uninterrupted service operation, which is the service's key characteristics from the clients' perspective.  The monitoring is done by gathering telemetry from the components in the forms of logs, execution traces, metrics, etc. Hereon, for the sake of brevity, we will collectively refer to these telemetry records as \textit{logs}. 

Generally, logs are produced by every component and are collected continuously (24/7). The majority of logs are textual; some are in machine-readable formats. The logs' sources vary (e.g., hardware logs, operating system logs, and application logs), which reflect the heterogeneity of components. Increasing complexity, velocity, variety, and volume of logs generated by Cloud platforms components makes the analysis of such logs a Big Data challenge. The challenges and solutions for operational log analysis are discussed in~\cite{miranskyy2016operational}. The challenges related to Cloud monitoring are explored in~\cite{pourmajidi2017challenges}; a Cloud monitoring solution that uses IBM Cloud services to monitor the IBM Cloud platform is proposed in~\cite{pourmajidi2019dogfooding}.

In recent years, advanced and automated monitoring systems have been implemented in data centres to monitor Cloud components' health. However, most existing monitoring systems still depend on statistics and heuristics from the observed health indicators  centred on resource usage thresholds~\cite{pourmajidi2017challenges, pourmajidi2019dogfooding}. In a Cloud environment, such anomaly detection techniques are less effective due to the Cloud platforms' scale and the rapidly changing nature of the workloads executed on these platforms~\cite{pourmajidi2017challenges}. The reduction of efficiency is typically manifested by numerous false alarms that overwhelm the operational team~\cite{pourmajidi2019dogfooding}.

Our software under study is the IBM Cloud Console, hereon the \textit{Console}, which provides the web frontend to the IBM Cloud. The IBM Cloud is an open and secure public Cloud providing enterprise-class IaaS and PaaS capabilities in 60+ data centres distributed around the globe~\cite{IBMCloud78:online}. The IBM Cloud also provides over 170 managed services covering data, AI, IoT, security, blockchain, and more. The Console is one of the modules of the IBM Cloud Platform, hereon the \textit{Platform}, which encompasses core functionality tying together all parts of the IBM Cloud, such as identity and access management, billing, search and tagging, and product catalog. The Platform evolves using DevOps principles, i.e., it is developed, maintained, and operated by a loosely-coupled geo-distributed IBM team. 

The Console is an exciting product to study in this ecosystem because its nearly 100 microservices (each geographically deployed across 10 data centres) depend on APIs from nearly every part of the Platform as well as the rest of the IBM Cloud. The Console's telemetry data for all of the microservices consists of millions of data points representing every inbound and outbound HTTP request each day. The Console then acts as a ``canary in the mineshaft" providing great insights into how well the entire IBM Cloud is working at any given point in time. 

The DevOps principle enables rapid response: if required, a developer can be brought to help with the triage and root causes analysis of service degradation. On the upside, this leads to speed up of identifying a root cause and higher reliability and availability of the Cloud services. This also acts as a strong motivating factor to deliver changes in small increments and thoroughly test the changes in a staging environment: nobody wants to be on a service call in the middle of the night. There is also a downside: the more developers spend diagnosing operational issues, the less time they have remaining to deliver new functionality.

As with any complex System-of-Systems, it is impossible to guarantee trouble-free operation of the Platform. There are numerous sources of trouble, e.g., system misconfiguration, software failure due to a hardware failure or a software bug that escaped into the production environment, or performance degradation due to the processing queue's saturation.

Thus, we want to detect anomalous behaviour before it escalates to severe service degradation or an outage that will impact the client. Often, if caught early, a problematic component can be fixed before it starts affecting other inter-related components and, in turn, clients. Early detection is also vitally important to meeting the Service Level Agreements (SLAs): relatively low availability of 99.99\% translates into one minute of downtime per week~\cite{piedad2001high}. This implies that the majority of problems must be resolved before they affect a client. 

As discussed above, the heuristic-based indicators often generate a large number of false alerts. Theoretically, the DevOps team may manually monitor the streams of logs emitted by thousands of the Platform components to detect problems. However, to make it efficient, this task requires a lot of human resources, making the approach economically infeasible. Thus, we need to create an automatic monitoring solution that can analyze the telemetry from thousands of components in near-real-time and alert the DevOps team about the abnormalities. 

This paper describes a machine-learning-based anomaly detector used for proactive detection of problems in the Platform's components. The detector can capture anomalies up to 20 minutes earlier than the previously existing one. We discuss the detector's architecture and design and share lessons learnt from integrating the detector into the Platform's DevOps pipeline.  We hope that the lessons learned from our work will inspire other researchers and practitioners to build their own reliable and scalable anomaly detection solutions for their Cloud systems.

The rest of this paper is structured as follows. We describe our challenges in Section~\ref{sec:challenges}. Then we present the architecture of our solution (addressing our challenges) in Section~\ref{sec:proposed_solution}, discuss its implementation in Section~\ref{sec:technical}, and recap the key aspects in Section~\ref{sec:developed_tools_processes}. Next, we report insights and best practices in Section~\ref{sec:insights_best_ractices} and provide conclusions in Section~\ref{sec:conclusions}.

\section{Our challenges}\label{sec:challenges}

Maintaining an adequate response time of IBM Cloud Platform/Console application components is crucial for the DevOps teams. An increase in response time might indicate a malfunctioning of the Platform components, leading to service interruption. Thus, proper monitoring is essential to ensure service quality and keep up with the SLA. Though the current monitoring system encompasses the components to collect data, the statistics-based anomaly detectors generate too many alerts~\cite{pourmajidi2017challenges,pourmajidi2019dogfooding}. The flooding of alerts (many of which are false) makes it impractical to rely and work on those. Furthermore, in several occurrences, the current monitoring system was unable to detect the deviation and failed to raise an alert. Thus we need to build a streaming anomaly detector that maps anomalies to the true problems and filters out the false ones.

\subsection{Monitoring system}
An advanced monitoring system has been deployed by the Console to observe the Platform components' health. Each Console microservice is written with Node.js and embeds an agent that emits data for each inbound and outbound HTTP request to a message queue. In the current system, the message queue is provided by the IBM IoT Platform, which uses the Message Queuing Telemetry Transport (MQTT) protocol. Hereon, we will refer to the IBM IoT Platform as Message Queue (PMQ).

The PMQ receives the logs from all of the data centres in which Console microservices reside. The log records are passed in the payload of PMQ messages. The log records are in JSON format, making it easy to parse them. Listing \ref{lst:json-example} shows an example of a log record capturing response time from a particular microservice. In addition to the response time (captured in the \code{responseTime} field), the records contain information about the microservice (shown in \code{appName}), URL (shown in \code{url}), host providing the API being called (shown in \code{target-endpoint}), HTTP response code (shown in \code{statusCode}), and HTTP method used in a given interaction (shown in \code{method}).

We require an additional data transformation sub-pipe, which enables our ML model by transforming and aggregating the original MQTT pipeline's raw data. Furthermore, we want to divert these transformed observations into persistent storage as MQTT is designed to ``fire and forget.''

\begin{listing}
\small
\begin{minted}[frame=single,
               framesep=3mm,
               linenos=true,
               xleftmargin=21pt,
               tabsize=4]{js}
{
  "eventType": "performanceMetics",
  "appName": "catalog",
  "url": "/abc/v1/accounts/1e...",
  "groupedUrl": "/coe/v2/...",
  "urlCategory": "URL-CATEGORY-1",
  "targetEndpoint": "x.somecloud.com",
  "timestamp": 1601824467164,
  "responseTime": 254.84029,
  "statusCode": 200,
  "method": "GET"
}
\end{minted}
\caption{Example GET request in JSON format.} 
\label{lst:json-example}
\end{listing}

\subsection{Big Data characteristics}\label{sec:Big data characteristics}
The \textit{velocity} of log records is high: each data centre emits thousands of records per minute. This high velocity of logs translates into large \textit{volume} of log records that have to be stored. While, by design, the log records are machine-readable, they have different formats, hence the \textit{variety} of the data. We have reviewed  Cloud-generated logs exhibits Big Data characteristics~\cite{Mockus2014EBD,jin2015significance,miranskyy2016operational,hashem2015rise,lemoudden2015managing,pourmajidi2017challenges}. These characteristics impose design and implementation challenges for Cloud monitoring platforms. To address these challenges, we describe some of the desired attributes of a successful Cloud monitoring platform in Section~\ref{subsec:Desired Characteristics}.

\subsection{Real-time analysis}
There are mainly two design options for building a learning model (based on the data usages in the modelling pipeline \cite{hoque18icsa, bisong2019batch}). The first one is batch learning, also known as offline learning, while the second one is online learning. 
In batch learning, the machine learning model is built and trained using the entire available dataset at rest (in hand) at a certain point in time. After sufficient training and tuning with the test dataset, 
the model is shipped to the production, and learning ends there. 
Batch learning becomes inappropriate to deal with situations where data gets generated continuously from the source.  
In online learning, data flows in streams into the learning algorithm and updates the model. This data flow can be seen as individual sample points in the dataset or mini-batches \cite{hoque18icsa, bisong2019batch}. This model is ideal in situations where we need to use real-time data samples to build a prediction model. The perfect instance of these cases is Console components monitoring or stock market prediction, where data are generated continuously in time.

The Cloud monitoring process includes the strategies and practices for real-time observing, tracking, analyzing, and reviewing these resources and services, usually in a much higher magnitude than typical on-premises data centres. As the resources and services are centralized, IBM usually is responsible for deploying, providing, and maintaining their digital assets' performance as per the agreed SLA. Monitoring is an essential component and plays a vital role in the Cloud platform offerings, and Cloud Service Providers (CSPs) should continuously monitor the health of their products to ensure
Quality of Service (QoS). The quality dimensions of Cloud services are availability, reliability, and performance~\cite{pourmajidi2017challenges}. A monitoring platform demands complete and thorough visibility of the networks, devices, and services in real-time or near real-time to live up to the expectations.

\subsection{Pre-defined thresholds' inapplicability} \label{threshold-based-detector}
Existing health monitoring systems rely on statistics and heuristics based on predefined thresholds. These thresholds can not be used to mark Cloud-generated metrics as ``acceptable'' or ``anomalous''~\cite{pourmajidi2017challenges,pourmajidi2019dogfooding}. The variety and the random nature of today's application requirements make such anomaly detection techniques inadequate. Time series events often contain nonlinear trends that impact the accuracy of the statistical forecasting models. One needs to consider seasonality and norm for each metric.  E.g., while 300~ms is not an acceptable response time for an API call, it may be acceptable for a database query. 

\subsection{Data modelling}
To detect such complex abnormalities, one needs to look at the metrics collected from multiple hardware (e.g., CPU and disk utilization) and software (e.g., virtual machines and database engines) components holistically. In other words, one needs to aggregate these metrics with one anomaly detection model rather than building individual models for each component, which has been done by multiple authors~\cite{DBLP:conf/noms/WangTSR10, DBLP:journals/corr/HochenbaumVK17, DBLP:journals/isci/BhuyanBK16}. This approach leads to an increase of computational complexity for calibrating and training the models, as the dimensionality of input increases with the number of components. A single data centre contains thousands of components, each of which may emit multiple types of health metrics. Thus, the data under study are high-dimensional.

Based on our preliminary experiments with empirical data, one may leverage statistical tools, such as the Autoregressive Integrated Moving Average model with exogenous multi-seasonal patterns (x-SARIMA)~\cite{box2013time}, to detect the anomalies. However, time series from other sources (e.g., CPU and memory utilization) exhibit properties (e.g., abrupt jumps or flat areas) that do not have an obvious seasonality pattern, which makes x-SARIMA struggle with the Console data.  For such cases, one may employ Mean-Shift models (e.g., Changepoint~\cite{sturludottir2017detection} and Breakout~\cite{james2016leveraging} techniques). In principle, x-SARIMA can be extended to perform online learning and handle multivariate time series~\cite{Brockwell2016}.  While one can combat these issues by preprocessing the time series, this would require manual intervention, which becomes uneconomical.

To summarize, we dealt with an existing monitoring and notification system that suffered from flooding of false alerts. The high dimensionality and non-linearity of the data made it difficult to use statistical models.

\section{Proposed solution: architecture}\label{sec:proposed_solution}
To address the challenges mentioned above, we designed a layered, microservice-based architecture that constructs a scalable, reliable data collection pipeline. The pipeline can collect/pull data from various sources, e.g., message queues and databases. Once the collected data are validated and normalized, they are sent to the analytics component that marks observations as normal or anomalous. The anomalies are reported to  the Console DevOps team for feedback and validation. The processing pipeline is designed to be not dependent on the nature of the collected data, and the same solution can be used to assess and analyze various software, hardware, and network metrics. 

In this section, we provide technical details related to the proposed solution. We elaborate on the Cloud monitoring solution's   desired characteristics in Section~\ref{subsec:Desired Characteristics} and provide details of the layered  and serverless architectures in Sections~\ref{subsec:Layered architecture} and \ref{subsec:Serverless operation}, respectively. In Section~\ref{subsec:Data Sources: pulling and pushing}, we provide additional details on various data source that our solution can support and conclude this section by providing implementation details in Section~\ref{subsec:Implementation}.

\subsection{Desired characteristics}\label{subsec:Desired Characteristics}
 
Here we list the desirable key characteristics of a  large-scale Cloud monitoring system~\cite{pourmajidi2019dogfooding}. 

\textit{Scalability}: A system that can monitor several metrics for each of the components distributed among several data centres needs to be scalable. Furthermore, this system has to be able to support Cloud-generated logs that exhibit Big Data characteristics. These characteristics were reviewed in Section~\ref{sec:Big data characteristics}.

\textit{Elasticity}: A system should automatically scale up or down with the intensity of the incoming log records. 

\textit{Reliability and Stability}: A monitoring system is mainly implemented to ensure that all other components of a Cloud platform are operating normally. As a collection of data and monitoring of a Cloud platform are continuous activities and require 24/7 operation, the system  should be reliable and resilient to component failure. 

\textit{Capacity}: The scale of generated logs requires a monitoring system with elastic capacity. That is, the size of the collected logs continues to grow, and so does the required space to store them. 

\textit{Support of various log formats}: A Cloud platform consists of several different types of components, and the logs and their formats are heterogeneous. Thus, a monitoring system should be able to collect and process various types of logs. 

\textit{Interconnection Feasibility}: As Cloud providers continue to add new services, the existing monitoring system should keep up with new demands. 

\subsection{Layered architecture}\label{subsec:Layered architecture}

Our solution, shown in Figure~\ref{fig:architecture}, is based on the 7-layered architecture for processing online and offline data\footnote{A detailed description of the 7-layered architecture is given in \cite{hoque2018architecture,hoque18icsa}.}. 
This architecture allows us to use microservices and publish-subscribe (pub-sub) architecture pattern and offers a good balance between scalability and maintainability due to high cohesion and low coupling of the solution. Furthermore, asynchronous communication between the layers makes the layered architecture a building block for a general architecture for processing any types of streaming data. 

For this work, we have adjusted the 7-layered model by combining Layers 1 and 3 (converter and splitter) as one composite layer and Layers 5 and 7 (aggregator and modeller) as another one. Hence, our solution has two composite layers.

In the 7-layered architecture, the pub-sub (implemented in Layers 2, 4, and 6) is used to establish communication between the odd layers. 
Once our codebase stabilized and we were confident in the high availability of the underlying services, we made a tradeoff decision to enable direct communication between the odd layers (removing pub-sub). Our decision simplified implementations and freed resources but reduced flexibility and reliability. 

The layered architecture's flexibility allows such modification as long as each layer has an atomic task and a unique business logic. Given that the incoming logs data were already in JSON format, we had minimal extract, transform, and load (ETL) efforts, and therefore merging of layers responsible for ETL was financially and technically feasible. To adhere to the layered architecture principles, we have kept the modules within a composite layer as isolated programming components so that their internal communicating can resemble the communication among real layers and pave the way for future decoupling practices.

\subsection{Serverless operation}\label{subsec:Serverless operation}

We choose serverless architecture as our primary computation infrastructure for this work. Serverless architecture is a Cloud architecture that provides more operational control to the CSPs. It is a highly scalable and event-driven architecture and only uses resources when a specific function or trigger occurs. The CSP usually handles the capacity planning and management tasks such as a server or cluster provisioning, patching, operating system maintenance for developers~\cite{CloudFun20:online}. Therefore, developers focus on the innovation and improvements of the applications and services, which eventually brings agility to application development. 

Among the several significant benefits, scalability and simplicity in deployment are prominent. We are outsourcing the responsibilities of infrastructure management, including servers and databases, to the CSPs. Thus, the developers can focus on building code for the application or services. The CSP also supports computation resources on-demand, which helps to achieve the capacity at pace. As we do not have to build and maintain a server from the ground-up, it reduces ownership cost. Additionally, as pay-as-you-go is the most common billing option in serverless architecture, the expenditure model is shifted from the capital-expenditures-based model to the operating-expenses-based in the IT ecosystem for companies.
Moreover, deploying and maintaining a data centre has an overhead for power, computation, cooling, and technical human resources to manage them. As CSPs manage those in a centralized manner, they can optimize the computation resources and involve quality human resources with the scale's edge. Eventually, it helps make the overall solution relatively eco-friendly by maintaining a minimum server and resource footprint.

\subsection{Data Sources: pulling and pushing}\label{subsec:Data Sources: pulling and pushing}

Irrespective of the type of databases that stores the Cloud logs, our platform needs to  pull (retrieve) data from such data sources. The \textit{pulling} happens either by using the database-provided APIs that expose an interface to retrieve data or through database-provided software development kits and libraries.  

Similarly, many Cloud monitoring tools can be integrated with other 3rd-party platforms by calling an API  exposed on the 3rd-party side. This \textit{pushing} mechanism is often predefined and requires little to no changes in the current structure of the monitoring platform. Hence, it is one of the most common practices of system integration. The first composite layer of our architecture is exposed to the outside world for data collection. A RESTful API~\cite{fielding2000architectural} is designed and implemented, and logs can be pushed to this API.

\subsection{Implementation}\label{subsec:Implementation}
\begin{figure*}[htbp]
\centerline{\includegraphics[width=0.9\textwidth]{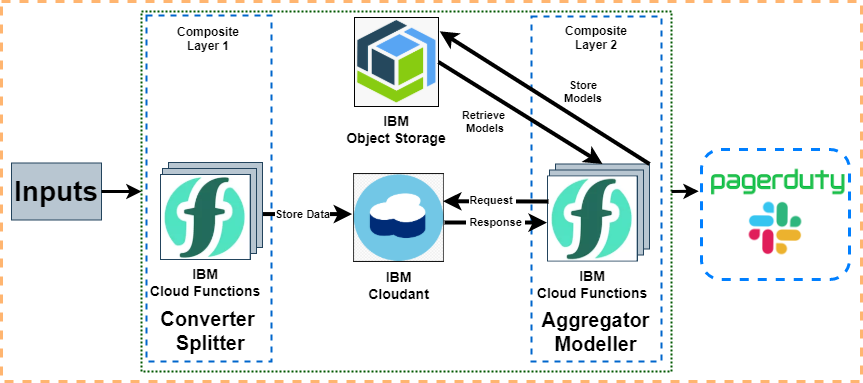}}
\caption{The layered microservice-based architecture of the proposed solution. The ``converter and splitter'' reside in the Composite Layer 1 of our architecture, the ``aggregator and modeller''~--- in the Composite Layer 2.  }
\label{fig:architecture}
\end{figure*}

In our  monitoring system, we leveraged public IBM Cloud services to implement the layered architecture. The relations between the services are represented graphically in Figure~\ref{fig:architecture}. The machine learning data collection and analysis pipeline is distributed over the two composite layers of one and two. Data wrangling and filtering happen in layer one, while the data analysis and anomaly detection happen in layer two.  

It is important to mention that while we have used IBM backed Cloud services, one can replace each one of these components with similar alternatives provided by different vendors. For instance, the composite layers one and two can be implemented using any any proprietary serverless computation infrastructure, such as AWS Lambda~\cite{AWSLambd42:online} and Google Cloud Functions~\cite{CloudFun61:online}, or open source ones, such as Apache OpenWhisk~\cite{ApacheOp33:online} and OpenFaaS~\cite{OpenFaaS:online}. The IBM Cloud Object Storage can be replaced by any proprietary object-based storage services, such as AWS S3~\cite{CloudObj90:online} and Microsoft Azure Blobs~\cite{AzureBlo53:online}, or a private deployment of an open-source service, such as Parse Server~\cite{ParseOpe49:online} and OpenIO~\cite{OpenIOHi69:online}. Similarly, IBM Cloudant can be replaced with another proprietary object storage database, such as AWS Dynamo DB ~\cite{AmazonDy75:online} and Google Cloud Firestore~\cite{Datastor73:online}, or a custom deployment of open-source database, such as Apache CouchDB~\cite{ApacheCo24:online} and MongoDB~\cite{Themostp23:online}. Additional details about the components of our platform are provided below.

\subsubsection{Publish-subscribe}

Although we did not include the pub-sub layers in the final product for this work, we initially implemented them by creating an instance of IBM Cloud Event Streams~\cite{EventStr87:online} (based on the Apache Kafka~\cite{ApacheKa3:online} software which implements pub-sub architectural pattern). The offering is fully managed, highly available, and is charged by the number of topics and the amount of data passing through them.

\subsubsection{Serverless Microservices}

We could have implemented microservices using managed container offering (namely, IBM Cloud Kubernetes Service~\cite{IBMCloud91:online}). However, we decided to leverage fully managed function-as-a-service (FaaS) offering of IBM Cloud Functions~\cite{CloudFun20:online} (which is implemented based on Apache OpenWhisk~\cite{ApacheOp33:online}). Given that the layered architecture is aimed at the Data Science pipeline, which is typically stateless, this was a natural decision. Solutions on FaaS scale up and down trivially, as the FaaS platform `fires up' individual instances of a Function to process incoming messages: the more messages are in the pipeline, the more Functions will be running concurrently. It was also attractive economically: FaaS are billed based on the amount of time a Function is running (proportional to the number of CPUs and memory requested to each Function); thus, we do not have to pay for idle resources.

We implement the Functions (from hereon, we will use the term `Function' and `microservice in the layered architecture' interchangeably) in Python. 

IBM Cloud Functions have multiple types of triggers to start a Function. Our solution receives incoming data via RESTful API~\cite{IBMAPIEx40:online}. IBM Cloud has readily available service to deploy the API and authorize access to it~\cite{IBMAPIEx60:online}. Triggers are set up to call a corresponding Function in the Composite Layer 1 for every POST request coming through the API.

Low-dimensional models with a few hundred dimensions can be fully trained in the IBM Cloud Functions serverless environment. However, as the model's complexity grows and dimensionality increases to thousands of features, the training time may exceed the 10-minutes timeout limit of the IBM Cloud Function. In that case, the model can be moved to a novel serverless IBM Cloud service called Cloud Code Engine~\cite{AboutIBM67:online} (which has a 2-hours timeout for a job) or trained in a Cloud-hosted virtual machine (VM) that is instantiated on demand. Cloud Code Engine was introduced in the Fall of 2020, long after we started this project. Thus, our solution was to use  IBM Cloud Functions for online training and reporting; and a VM for periodic batch re-training (which happens asynchronously). We are planning to migrate the re-training to the Cloud Code Engine to make the solution completely serverless.

\subsubsection{Persistent storage}

For persistent storage, we use two services. The first one is a fully managed JSON document store IBM Cloudant~\cite{IBMCloud29:online}, which elastically scales throughput as well as storage. However, a single document, stored in Cloudant, cannot be larger than 1~MB~\cite{Limits67:online}, which implies that we cannot store large transformed data frames and trained models. For these items, we leverage IBM Cloud Object Storage~\cite{CloudObj83:online}, another fully-managed and scalable service.

\section{Proposed solution: notes on implementation}\label{sec:technical}

\subsection{Input Data handling}
\subsubsection{Data filtering}
We monitor the incoming traffic from the PMQ and filter the relevant telemetry records as per the DevOps teams' given specification and criterion required for our analysis. This refining process reduces the number of data points from thousands to approximately 200 per minute per data centre.

\subsubsection{Data aggregation}
The filtered relevant data points are cached and aggregated periodically. This is done to reduce the data volume and reduce the amount of noise in the inputs.

First, we aligned time intervals with astronomical time. Then we aggregate the data in 5-minute intervals (the interval's length is chosen based on the DevOps team's feedback). The aggregation functions are minimum, maximum, count, median, mean, standard deviation, and skewness. 

Finally, there might be some stale records, provided that the data arrives asynchronously. In that case, on those time intervals on which these data records come on, we re-run the anomaly detection.

\subsubsection{Feature selection}
The features are generated by grouping the response time over unique component id, REST method (e.g., GET, POST, or DELETE), REST status code returned by a service (e.g., 200, 401, or 502), and aggregate statistics defined above. We also add hourly, daily, weekly, and monthly seasonality features using standard trigonometric (sine and cosine) transformations~\cite{stolwijk1999studying}.

\subsubsection{Missing data handling} 

We have four types of missing values that can occur at any stage of training or testing:
First, nulls coming from aggregation functions like standard deviation and skewness. These measures require more than one observation in a group of observations. If there are not enough points to compute these statistics, they are replaced with zeroes. Second, in a given period, the absence of observations of certain features. These are replaced with zeroes too. Third, there are no observations (at all) in a given time interval. We keep all rows as zero, send a warning and make the prediction. Fourth, in the are no observations from the latest four time intervals --- send a warning and do not make a prediction. 

\subsubsection{Data normalization}

With this aggregation of the platform applications' collected response time data along with the feature groupings, the one-dimensional input data turned into multi-dimensional time series. Then we handled the missing data as discussed above. Finally, we have normalized the multi-dimensional datasets before training the models. The resulting matrix is normalized to the $[0,1]$ range using each feature's min and max value during the training. Sometimes this value may end up being outside of this range if a new observation is outside of the original min-max range during online learning. However, this did not affect the detector—this behaviour self-corrects during the next batch re-training session.

\subsubsection{Never-seen-before features processing} \label{unseen-feature}
Given the Cloud services' dynamic nature, new components and services are deployed frequently, resulting in new features. If such a feature is detected, we omit it from online training and send a warning indicating that a re-training (to include this feature) may be needed. While most of these warnings are harmless, some may indicate an underlying problem. For example, a never-seen-before 500 code (“the server encountered an unexpected condition which prevented it from fulfilling the request”) may indicate a defect in the service’s code. Thus, we prioritize warning messages based on the return codes. Finally, during the next batch re-training, these never-seen-before features are automatically added to the re-trained model, which disables the warnings.

\subsection{Anomaly detector model selection}
We are building a machine-learning-based scalable anomaly detector capable of detecting failures automatically in Cloud components. Generally, the time series analysis intends to gain insight into the dynamics behind time-ordered data. Anomaly detection is an old research area in many disciplines with a large body of research \cite{hawkins1980identification, hodge2004survey, zhang2017time, hochenbaum2017automatic}. Still, the seasonality and trending nature of Cloud infrastructure data limits many statistical techniques \cite{hochenbaum2017automatic}.

However, it is a non-trivial task to handle non-stationary multi-dimensional time series using classical statistical tools, as described in Section~\ref{threshold-based-detector}.  On the contrary, Artificial Neural Network (ANN)-based models may have nonlinear activation functions that can effectively extract nonlinear relationships in the data. They are suitable for time series predictions because of their characteristics of robustness, fault tolerance, and adaptive learning ability \cite{wang2011solar}. 

Therefore, we have explored  deep neural networks (DNNs), specifically recurrent-neural-network-based models~\cite{lipton2015critical} due to the suitability of handling sequential data. To analyze time series data and identify anomalies, we have tried Long Short-Term Memory (LSTM) \cite{hochreiter1997long} and Gated Recurrent Unit (GRU) \cite{cho2014learning} models. Eventually, we concentrated on GRU-based models. Based on our experience, of all the models we experimented with, they gave the strongest predictive power. 
To the best of our knowledge, we are the first to use the GRU-based autoencoder model incorporated with the likelihood function for anomaly detection in multi-dimensional Cloud components telemetry~\cite{islam2020anomaly, mohammads_thesis}. The works that are closest to ours are Numenta's Hierarchical Temporal Memory (HTM) \cite{ahmad2017unsupervised} based learning algorithm for error calculation in time series, and National Aeronautics and Space Administration (NASA)'s  LSTM networks-based anomaly detectors \cite{hundman2018detecting}. However, these works are complementary to ours. Although both used their models along with the likelihood function as we did, the model structures are different (HTM and LSTM vs. GRU). Moreover, HTM processes only one-dimensional data, while NASA's multi-dimensional LSTM models are tested in a domain different from ours (i.e.,  ``rocket science'' vs.  ``Cloud components'').

Our GRU-based autoencoder gets trained in a self-supervised manner. To enhance the detection's performance, we have computed the likelihood of anomalies from the reconstruction error scores distribution. 

To design a dimension-independent neural network model to detect outliers in a Cloud platform's components, we have explored and calibrated the hyper-parameters controlling GRU architecture. We evaluated the performance against a publicly available single-dimensional labelled dataset and multi-dimensional datasets of the Console. Our model yields adequate results for the one-dimensional Numenta Anomaly Benchmark (NAB) benchmark dataset with a score of 59.8 on the Standard Profile, which is currently the third-best score \cite{islam2020anomaly}. But, we have not performed a quantitative comparison for the multi-dimensional telemetry due to the lack of a standard benchmark in this field. 
Therefore, we relied on the IBM Platform DevOps teams' anecdotal feedback on the anomalies detected by our model in multi-dimensional datasets. Scoring a satisfactory result for the diverse dataset is an indication of our model's generalization capability. 

\subsection{Model calibration}

We used the GRU-based autoencoder to get the reconstruction error; then, we initially marked the anomalies based on the error values. We saw that the raw error values' predictive power is low against publicly available benchmark datasets (NAB score of $< 40$). Therefore, the likelihood transformation was in order \cite{islam2020anomaly}. We use the anomaly likelihood measurement approach introduced by Ahmad et al. \cite{ahmad2017unsupervised}. It maintains a window of recent error  values and incrementally processes raw errors. Historical errors are modelled as a normal rolling distribution of the last $W$ points window at each step $t$. We measure the probability of anomaly by analyzing the distribution of the autoencoder's reconstruction error scores that significantly increased the anomaly detection capability and corresponding scores for the same benchmark datasets. 
Moreover, as per our design, we can tune the anomaly reporting by changing the anomaly likelihood threshold rather than the direct reconstruction error level. This gives better control over the model's output, for example, the reported false positive amounts, which is a key pain point of the Console DevOps team.

We have explored the parameters that govern the GRU architecture that yield the best predictive power for our time series. Next, we are looking for an adequate input data window size: how many observations in the time series are required to capture the intricate patterns. 
Then, we have used a grid search approach for the hyper-parameter tuning of the GRU-based model. 
We also make sure the adjustments and parameter tuning does not slow down the computation significantly. 

\subsubsection{Initial training of the model}
We have made a combination of offline and online training for our model. For the initial training, we use the offline training method.  We have selected a reference period to make the autoencoder understand the normal situation within the input data stream. We calculate the reconstruction error in terms of Mean Squared Error for the training dataset in the autoencoder model. 
We have taken the first $9,000$ data points from input data for performing the initial batch training. This ensures that we are using nearly a month worth of data ($12 \times 24 \times 30 = 8,640 \approx 9,000$ data points based on a 5-minute data aggregation interval). From our experience, it helps the model to identify hourly, daily and weekly seasonal patterns. 
For smaller datasets where this amount is infeasible, we recommend using at least 15\% of dataset length for initial training of the model. 
The rest of the dataset is used for testing/evaluating and online training of the model.

\subsubsection{Evaluation and online training}
After completing the initial training, the model is deployed into a Cloud Function and invoked every five minutes. Once invoked, it evaluates new telemetry data to detect anomalies and updated the model (using an online training approach). It takes less than 20 seconds to perform this evaluation. If the data are deemed anomalous, a detailed report (discussed in Section~\ref{sec:presentation}) is generated and sent to the DevOps team. It takes 20 to 120 seconds to create the report (the duration depends on the number of highlighted anomalous features). Our detector can report anomalies up to 20 minutes earlier than the previous monitoring tools based on the DevOps team feedback.

\subsubsection{Periodic re-training}
We perform a periodic automatic batch re-training every six hours to incorporate the never-seen-before features (explained in Section~\ref{unseen-feature})  into the re-trained model. Thus, the model remains up to date concerning the input data changes through the offline batch re-training. The re-training happens asynchronously with the evaluation process and does not affect it.

\subsection{Reporting}\label{sec:presentation}

The detector sends out the alerts to the appropriate DevOps team using Slack channels. If two or more consecutive alerts appear within 10 minutes, the posts go to the same thread to reduce clutter. The Slack message contains a high-level summary of the alert and the count of anomalous features grouped by the Platform's component groups. It also includes a hyperlink to the detailed report. The detailed report contains a textual description of the anomalous features and the graphical representation of these features.

\begin{figure*}[htbp]
\centerline{\includegraphics[width=0.95\textwidth]{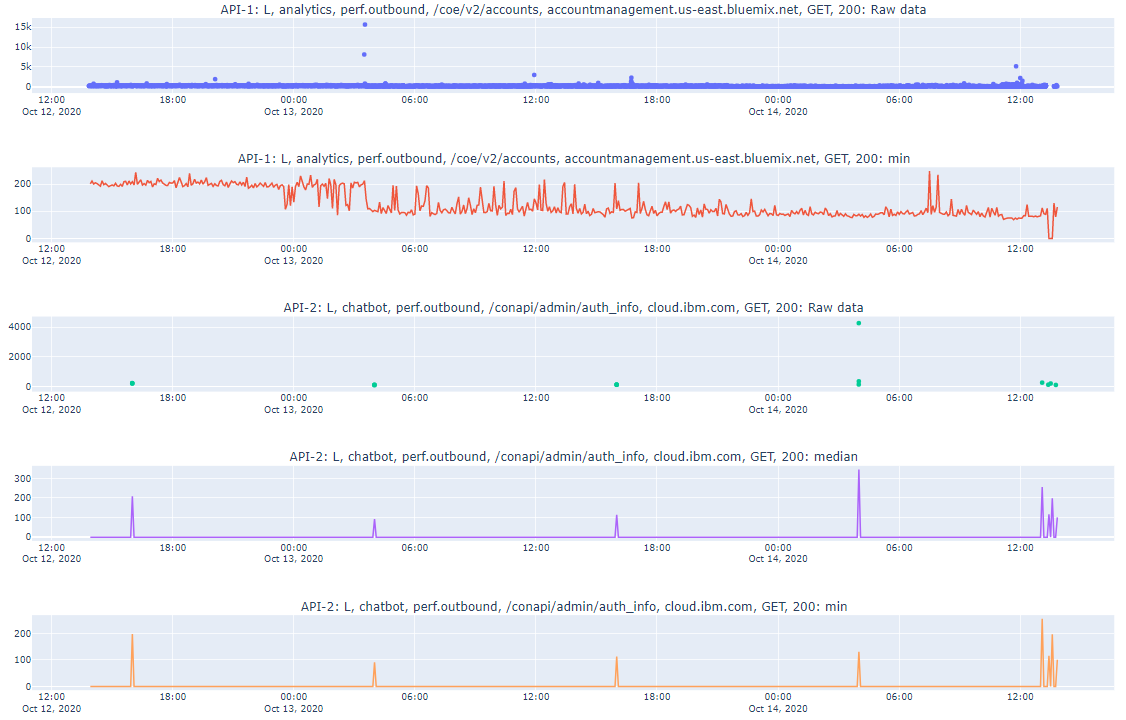}}
\caption{A graphical report received by the DevOps team. The report shows the anomalous aggregated features along with the raw data points for the last 48 hours.  The reported anomalous behaviour is shown in the last one hour of data on the plots' right-hand side. The analyst uses the remaining historical data to compare the features' historical behaviour with the recent anomalous one.
The top two panes show anomalous behaviour for API-1, the remaining ones for API-2. In API-1, we can see a gap in the data arrival, which has not happened in the past. In the API-2 case, the calls to the API happened periodically in the past but always resulted in a single ``spike''. But for the anomalous period, we see a continuous request pattern.}
\label{fig:all-plots}
\end{figure*}

An example of such a graph is shown in Figure~\ref{fig:all-plots}.  Note that anomalous behaviour does not always map to a problem. Rather, it shows that the model did not see such behaviour in the training data. We are training the model on the latest 9000 data points (approximate one month of data), which guarantees that reported anomalies are rare indeed.

\subsubsection{Feedback mechanism}
\begin{figure}[th]
    \centering
    \includegraphics[width=0.8\columnwidth]{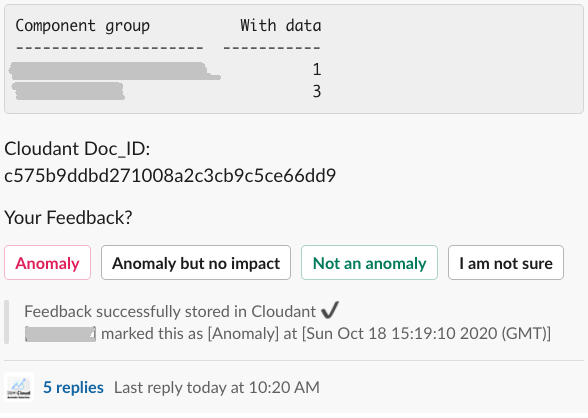}
    \caption{Slack alert with feedback mechanism by the end-users. This alert was considered ``Anomalous'' (impacting a customer) by the DevOps team member. The alert also shows a link to a thread of related alerts mentioned in Section~\ref{sec:presentation}.}
    \label{fig:slack-alart}
\end{figure}

As discussed above, anomalies do not always translate into problems, but only the DevOps team, can verify this.
Thus, we include a mechanism for the direct feedback from the DevOps teams for the reported anomalies in the Slack channel. They have four options to choose from, namely
    \begin{enumerate*}
        \item  ``Anomaly'' which impacts a client,
        \item  ``Anomaly but no impact'',
        \item  ``Not an anomaly'', and
        \item  ``I am not sure''.
    \end{enumerate*}
Multiple submissions are allowed. We keep track of all the clicks by storing those in the IBM Cloudant database and presenting the latest feedback and the feedback originator's name. An example of a message with feedback is shown in Figure~\ref{fig:slack-alart}. The models are then improved based on the received feedback.

\section{Proposed solution: recap}\label{sec:developed_tools_processes}
We have designed and deployed an ML-based streaming anomaly detector solution for Console components' multi-dimensional time series on the IBM Cloud service infrastructure. We used the PMQ to get the logs and utilized IBM Cloud Functions as the runtime environment (which extracts telemetry data from the logs, performs anomaly detection, and sends reports). 

Our tuned model can detect anomalies for apparent changes in the data set and quickly adjust to the new normal trend—the sensitivity (level of detection) can be controlled by choosing the appropriate likelihood threshold. The detector sends out the alerts to relevant DevOps teams using the Slack channel and collects their feedback with a button click. The feedback is then used to adjust the model.

We design our anomaly detector to be generic and capable of working with multi-dimensional time series of the different nature. Above, we discussed detection anomalies in the performance-related telemetry. However, we have also tested it on other types of data. An example of detecting anomalies in the  CPU, memory, and file system usage data is given in Appendix~\ref{sec:vm}.

\section{Insights and best practices that emerged}\label{sec:insights_best_ractices}

\subsection{Insights}
The Console depends on many external services. If the alerts are coming from microservices interacting with the external APIs, it may hint that the problem is external, and the root cause of the issue could be beyond the Console's ecosystem. In the identification of such an event, a shortlist of possible defective components helps. The granularity at the level of specific return codes and REST methods is significant, as it helps pinpoint the problem's root cause. Analysis of anomalies from multiple data centres helps to understand whether the issue is localized, i.e., confined to a specific data centre and can be mitigated by redirecting requests to another data centre, or the problem is a global one where a failover will not help.

Anomaly detection further helps identify Denial of Service attacks -- as they are easily detectable by anomalies related to spikes in the count statistics. It is worth mentioning that the attack may not always necessarily be malicious and coming from the outside. For example, it can be a mistake of a tester who executes a load test in the production rather than the staging environment. 

The 20 minutes heads up offered by the model may not sound like much. However, to put things into perspective, SLA with 99.999\% uptime allows approximately five minutes of downtime per year~\cite{piedad2001high}. Thus, 20 minutes of extra time gives DevOps team an extra  ``wiggle room'' that is appreciated.

\subsection{Best practices}
\subsubsection{Architecture and Implementation}
Serverless architecture enables autoscaling. This helps cope with the changing patterns in the log records arrival. 

NoSQL database as a service (DBaaS) document-centric database simplifies the ETL of our JSON-based log records. The DBaaS enables seamless adaptation to the elasticity of the incoming traffic.

\subsubsection{Modelling and workload}
One-class models are more suitable than two-class models for anomaly detection in the Cloud telemetry based on our experience. Real data are not labelled, and practitioners do not have time to provide detailed feedback/labelling. Moreover, sometimes they may not be aware of the problem in the first place. Our GRU-based autoencoder seems to be a perfect match for the task at hand.

Make sure that the workload on which the anomaly detector is tested resembles the workload of the environment that the detector will monitor in production. We initially used the logs coming from the Platform's staging environment for testing the anomaly detector, which ended up being a bad idea. The Platform's staging environment workload was not representative of its production environment's workload, capturing unusual patterns that may never appear in the production. Thus, we had to ensure that the detector was tested on the telemetry captured in the  Platform's production environment logs.

\subsubsection{Re-training}
Never-seen-before features have to be filtered out before passing them to the model, which requires a bit of extra coding. However, these features may provide additional diagnostic information and should be reported (as discussed in Section~\ref{unseen-feature}).

Thus, while we have online re-training, we still need to do batch re-training periodically, as new features (coming from newly deployed services and APIs) frequently appear in the Cloud environment. How often to do this re-training? This depends on the available computing resources and the necessity to get the DNN-based detection up to speed. We found that the 6-hour interval was adequate for our needs.

\subsubsection{Generalization} 
The solution, learning process, and best practices can be used with other data types, as long as they can be mapped to a numerical scale.

\section{Conclusion}\label{sec:conclusions}
In this paper, we investigated the challenges that the IBM DevOps team was facing, i.e., finding a reliable and scalable monitoring system for IBM Cloud Platform, which contains thousands of components. We proposed to solve the challenge by designing and implementing an automated real-time DNN-based anomaly detector, which alerts the DevOps team via Slack. The solution is monitoring the production environment of the Console of the IBM Cloud Platform for the last year. It detects complex anomalies in multi-dimensional time series up to 20 minutes earlier than the previous monitoring solution and significantly lowered the false alarm rate.

This work brings us closer to creating generic anomaly detectors and is, therefore, of interest to the academic community. It is also of interest to practitioners, as our findings, insights, and best practices can be leveraged in building production-grade anomaly detector in other fields.

\appendix

\section{Anomaly detection in the VM telemetry }\label{sec:vm}
\begin{figure*}[htbp]
\centerline{\includegraphics[width=0.95\textwidth]{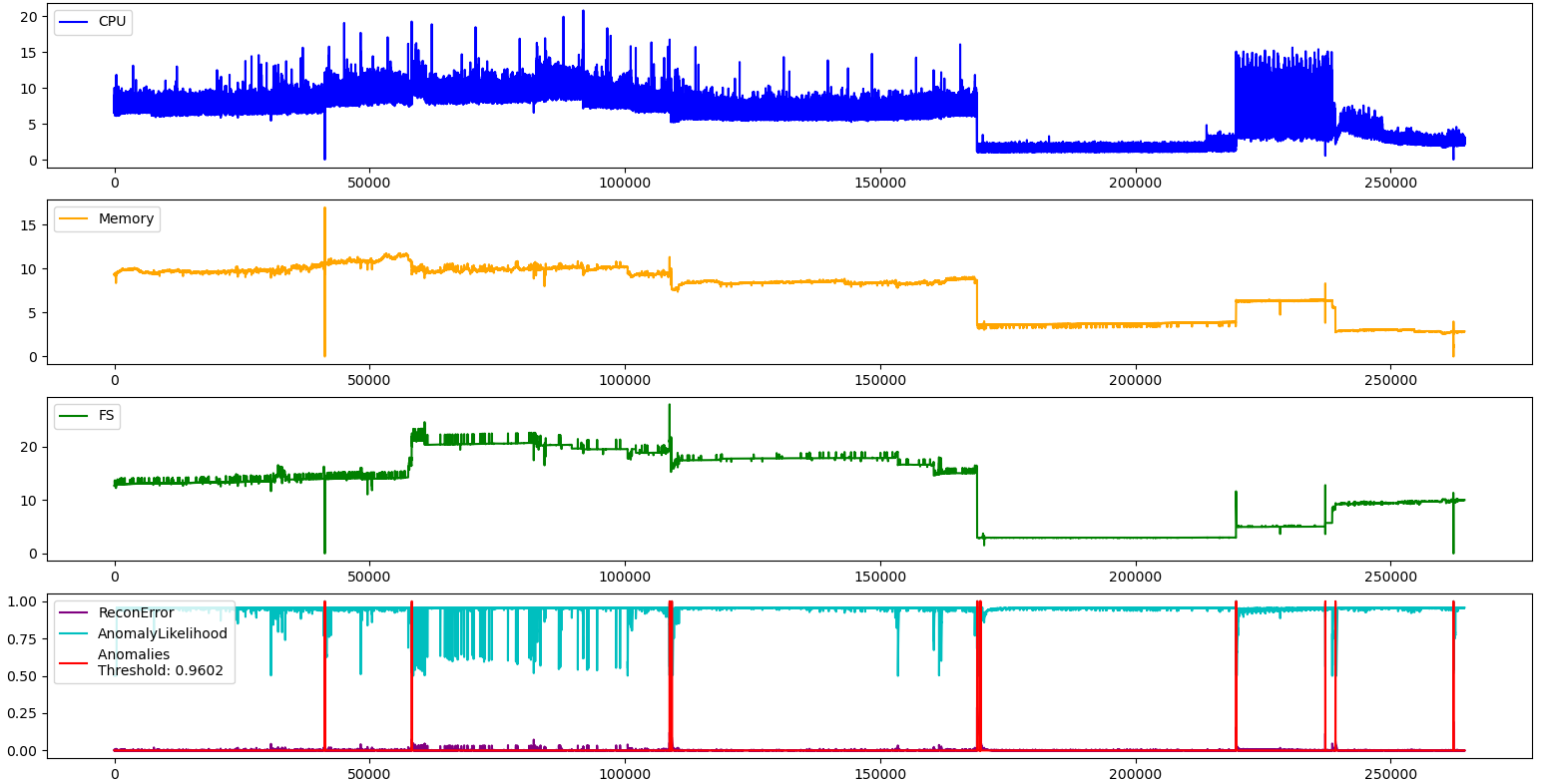}}
\caption{An example of anomaly detection in the telemetry of a Cloud-hosted VM. It shows a 1-month subset of data (265,200 points). Timestamps marked with an index on the $x$-axis for better readability. The top pane indicates CPU usage  percentage; the second-top pane shows memory usage percentage; the third-top pane --- file system usage percentage. The bottom pane shows reconstruction error (in purple), likelihood score (in cyan), and anomalies with the likelihood threshold of 0.9602 (in red). Red spikes correspond to the anomalies listed in Table~\ref{table:db2-reported-anomalies}. The third red thick spike from the left represents the Anomaly-03 and Anomaly-04; the fourth spike depicts Anomaly-05 and Anomaly-06 from Table~\ref{table:db2-reported-anomalies}.}
\label{fig:db2-anomaly}
\end{figure*}

Anomaly detection for a VM deployed in the IBM Cloud is shown in Figure~\ref{fig:db2-anomaly}. The telemetry was collected at 10-second intervals for 145 days, gathering 1,259,340 data points in total. An analysis of a subset of these data (for 1-month, 265,200 points) is shown in Table~\ref{table:db2-reported-anomalies} and Figure~\ref{fig:db2-anomaly}.

\begin{table}[ht]
\centering
\caption{Database server telemetry: Here we summarize the anomalies flagged by the model along with their duration for the 1-month  subset of data as indicated in Figure~\ref{fig:db2-anomaly}. The anomaly detector marks the start and endpoint indices, and anomalies can be matched with the red spikes in Figure~\ref{fig:db2-anomaly}.}
\begin{tabular}{ lrrr}
    \toprule
    Anomaly	& Start Point Index & End Point Index& 	Range\\
    \midrule
    Anomaly-01 & 41192 & 41228 & 37\\
    Anomaly-02 & 58234 & 58254 & 21\\
    Anomaly-03 & 108741 & 108747 & 7\\
    Anomaly-04 & 109163 & 109205 & 43\\
    Anomaly-05 & 168968 & 169059 & 92\\
    Anomaly-06 & 169527 & 169669 & 110\\
    Anomaly-07 & 219688 & 219745 & 58\\
    Anomaly-08 & 237183 & 237187 & 5\\
    Anomaly-09 & 239159 & 239178 & 20\\
    Anomaly-10 & 262272 & 262297 & 26\\
    \bottomrule
\end{tabular}
\label{table:db2-reported-anomalies}
\end{table}

Operation team confirmed if a given point-range was unusual or not. Thus, we relied on their feedback to place a reported anomaly into the true positives ($TP$), true negative ($TN$), false positive ($FP$), or false negative ($FN$) groups. The resulting confusion matrix is shown in Table~\ref{table:nD-confusionMatrix}. Based on this matrix, we have calculated the evaluation metrics shown in Table~\ref{table:nD-evaluation-scores}.

\begin{table}[ht]
\centering
\caption{Confusion matrix: reported result by the model; $TP$ and $TN$ are based on DevOps team's feedback. The reported anomaly points are given in Table~\ref{table:db2-reported-anomalies} and shown in the Figure~\ref{fig:db2-anomaly} }

\begin{tabular}{ lrrr}
    \toprule
    & Predicted Anomaly & Predicted Normal & Total\\ 
    \midrule
    True Anomaly & ($TP$) 309 & ($FN$) 0 & \textit{309}\\ 
    True Normal & ($FP$) 110 & ($TN$) 264781 & \textit{264891}\\ 
    \textbf{Total} & \textit{419} & \textit{264781} & \textit{\textbf{265200}}\\
    \bottomrule
\end{tabular}

\label{table:nD-confusionMatrix}
\end{table}

Our model can detect anomalies in apparent changes to the data and adjust to the new normal trend. We compared the anecdotal feedback of the DevOps team against the anomalies reported by the model. The results are shown in Table~\ref{table:nD-evaluation-scores}. The DevOps team found our anomaly flagging better than that of the existing approaches (see~\cite[Section 4.2.2]{mohammads_thesis} for details of the analysis).

Our model has a high accuracy of 99.95\%. Note that accuracy is not a very good measure for anomaly detection in time-series data (see~\cite[Section 3.2.1]{mohammads_thesis}). Thus, let's look at additional metrics.

Precision indicates the percentages of real/true anomalies among the model's data points reported as anomalous. This measure does not consider the anomalies that the detector might not indicate, i.e. the $FN$. In general, high precision means lower false alarms. Based on the operation team's feedback, the model did not miss any such cases in the experiment. The precision of $\approx$~74\% was an improvement over the existing DevOps team approaches.

Recall specifies how well the model identifies all anomalies. It is the percentage of reported anomalies against total anomalies in the dataset. Usually, high recall means that the model is good at detecting outages/unusualities when they occur. Our model did quite well by not missing any anomalies during reporting.  

The F1-score is the harmonic mean of precision and recall,  combining precision and recalls into a single score.  We got the F1 score of $\approx$~0.85 for this dataset.

\begin{table}[t]
\centering
\caption{Model performance for the database server telemetry time series, based on Table~\ref{table:nD-confusionMatrix} data. }
\begin{tabular}{ lr }
    \toprule
    Evaluation Metric & 	Score \\
    \midrule
    Accuracy & 0.9995  \\
    Precision & 0.7374 \\
    Recall & 1.0000 \\
    F1-score & 0.8489\\
    \bottomrule
\end{tabular}
\label{table:nD-evaluation-scores}
\end{table}

\bibliographystyle{IEEEtran}
\bibliography{references} 

\begin{thebibliography}{10}
\providecommand{\url}[1]{#1}
\csname url@samestyle\endcsname
\providecommand{\newblock}{\relax}
\providecommand{\bibinfo}[2]{#2}
\providecommand{\BIBentrySTDinterwordspacing}{\spaceskip=0pt\relax}
\providecommand{\BIBentryALTinterwordstretchfactor}{4}
\providecommand{\BIBentryALTinterwordspacing}{\spaceskip=\fontdimen2\font plus
\BIBentryALTinterwordstretchfactor\fontdimen3\font minus
  \fontdimen4\font\relax}
\providecommand{\BIBforeignlanguage}[2]{{%
\expandafter\ifx\csname l@#1\endcsname\relax
\typeout{** WARNING: IEEEtran.bst: No hyphenation pattern has been}%
\typeout{** loaded for the language `#1'. Using the pattern for}%
\typeout{** the default language instead.}%
\else
\language=\csname l@#1\endcsname
\fi
#2}}
\providecommand{\BIBdecl}{\relax}
\BIBdecl

\bibitem{GartnerF38:online}
\BIBentryALTinterwordspacing
Gartner forecasts worldwide public cloud revenue to grow 17\% in 2020. Gartner.
  [Online]. Available:
  \url{https://www.gartner.com/en/newsroom/press-releases/2019-11-13-gartner-forecasts-worldwide-public-cloud-revenue-to-grow-17-percent-in-2020}
\BIBentrySTDinterwordspacing

\bibitem{miranskyy2016operational}
A.~Miranskyy, A.~Hamou-Lhadj, E.~Cialini, and A.~Larsson, ``Operational-log
  analysis for big data systems: Challenges and solutions,'' \emph{IEEE
  Software}, vol.~33, no.~2, pp. 52--59, 2016.

\bibitem{pourmajidi2017challenges}
W.~Pourmajidi, J.~Steinbacher, T.~Erwin, and A.~Miranskyy, ``On challenges of
  cloud monitoring,'' in \emph{Proceedings of the 27th Annual International
  Conference on Computer Science and Software Engineering}, ser. CASCON
  ’17.\hskip 1em plus 0.5em minus 0.4em\relax USA: IBM Corp., 2017, pp.
  259--265.

\bibitem{pourmajidi2019dogfooding}
W.~Pourmajidi, A.~Miranskyy, J.~Steinbacher, T.~Erwin, and D.~Godwin,
  ``Dogfooding: Using ibm cloud services to monitor ibm cloud infrastructure,''
  in \emph{Proc. of the 29th Int. Conf. on Computer Science and Software
  Engineering}, ser. CASCON ’19.\hskip 1em plus 0.5em minus 0.4em\relax USA:
  IBM Corp., 2019, pp. 344--353.

\bibitem{IBMCloud78:online}
\BIBentryALTinterwordspacing
{IBM Cloud Global Data Centers: IBM Cloud}. [Online]. Available:
  \url{https://www.ibm.com/cloud/data-centers/}
\BIBentrySTDinterwordspacing

\bibitem{piedad2001high}
F.~Piedad and M.~Hawkins, \emph{High availability: design, techniques, and
  processes}.\hskip 1em plus 0.5em minus 0.4em\relax Prentice Hall
  Professional, 2001.

\bibitem{Mockus2014EBD}
\BIBentryALTinterwordspacing
A.~Mockus, ``Engineering big data solutions,'' in \emph{Proceedings of the on
  Future of Software Engineering}, ser. FOSE 2014.\hskip 1em plus 0.5em minus
  0.4em\relax New York, NY, USA: ACM, 2014, pp. 85--99. [Online]. Available:
  \url{http://doi.acm.org/10.1145/2593882.2593889}
\BIBentrySTDinterwordspacing

\bibitem{jin2015significance}
X.~Jin, B.~W. Wah, X.~Cheng, and Y.~Wang, ``Significance and challenges of big
  data research,'' \emph{Big Data Research}, vol.~2, no.~2, pp. 59--64, 2015.

\bibitem{hashem2015rise}
I.~A.~T. Hashem, I.~Yaqoob, N.~B. Anuar, S.~Mokhtar, A.~Gani, and S.~U. Khan,
  ``The rise of “big data” on cloud computing: Review and open research
  issues,'' \emph{Information systems}, vol.~47, pp. 98--115, 2015.

\bibitem{lemoudden2015managing}
M.~Lemoudden and B.~El~Ouahidi, ``Managing cloud-generated logs using big data
  technologies,'' in \emph{2015 Int. Conf. on Wireless Networks and Mobile
  Communications (WINCOM)}.\hskip 1em plus 0.5em minus 0.4em\relax IEEE, 2015,
  pp. 1--7.

\bibitem{hoque18icsa}
S.~Hoque and A.~Miranskyy, ``Online and offline analysis of streaming data,''
  in \emph{2018 IEEE International Conference on Software Architecture
  Companion (ICSA-C)}, April 2018, pp. 68--71.

\bibitem{bisong2019batch}
E.~Bisong, ``Batch vs. online learning,'' in \emph{Building Machine Learning
  and Deep Learning Models on Google Cloud Platform}.\hskip 1em plus 0.5em
  minus 0.4em\relax Springer, 2019, pp. 199--201.

\bibitem{DBLP:conf/noms/WangTSR10}
\BIBentryALTinterwordspacing
C.~Wang, V.~Talwar, K.~Schwan, and P.~Ranganathan, ``Online detection of
  utility cloud anomalies using metric distributions,'' in \emph{{IEEE/IFIP}
  Network Operations and Management Symposium, {NOMS} 2010, 19-23 April 2010,
  Osaka, Japan}, Y.~Kiriha, L.~Z. Granville, D.~Medhi, T.~Tonouchi, and M.~Kim,
  Eds.\hskip 1em plus 0.5em minus 0.4em\relax {IEEE}, 2010, pp. 96--103.
  [Online]. Available: \url{https://doi.org/10.1109/NOMS.2010.5488443}
\BIBentrySTDinterwordspacing

\bibitem{DBLP:journals/corr/HochenbaumVK17}
\BIBentryALTinterwordspacing
J.~Hochenbaum, O.~S. Vallis, and A.~Kejariwal, ``Automatic anomaly detection in
  the cloud via statistical learning,'' \emph{CoRR}, vol. abs/1704.07706, 2017.
  [Online]. Available: \url{http://arxiv.org/abs/1704.07706}
\BIBentrySTDinterwordspacing

\bibitem{DBLP:journals/isci/BhuyanBK16}
\BIBentryALTinterwordspacing
M.~H. Bhuyan, D.~K. Bhattacharyya, and J.~K. Kalita, ``A multi-step
  outlier-based anomaly detection approach to network-wide traffic,''
  \emph{Inf. Sci.}, vol. 348, pp. 243--271, 2016. [Online]. Available:
  \url{https://doi.org/10.1016/j.ins.2016.02.023}
\BIBentrySTDinterwordspacing

\bibitem{box2013time}
G.~E.~P. Box, G.~M. Jenkins, and G.~C. Reinsel,
  \emph{\BIBforeignlanguage{English}{Time series analysis: forecasting and
  control}}, 4th~ed.\hskip 1em plus 0.5em minus 0.4em\relax Somerset: John
  Wiley, 2008.

\bibitem{sturludottir2017detection}
E.~Sturludottir, H.~Gunnlaugsdottir, O.~K. Nielsen, and G.~Stefansson,
  ``Detection of a changepoint, a mean-shift accompanied with a trend change,
  in short time-series with autocorrelation,'' \emph{Comm. in
  Statistics-Simulation and Computation}, vol.~46, no.~7, pp. 5808--5818, 2017.

\bibitem{james2016leveraging}
N.~A. James, A.~Kejariwal, and D.~S. Matteson, ``Leveraging cloud data to
  mitigate user experience from ‘breaking bad’,'' in \emph{2016 IEEE Int.
  Conf. on Big Data (Big Data)}.\hskip 1em plus 0.5em minus 0.4em\relax IEEE,
  2016, pp. 3499--3508.

\bibitem{Brockwell2016}
\BIBentryALTinterwordspacing
P.~J. Brockwell and R.~A. Davis, \emph{Multivariate Time Series}.\hskip 1em
  plus 0.5em minus 0.4em\relax Cham: Springer International Publishing, 2016,
  pp. 227--257. [Online]. Available:
  \url{https://doi.org/10.1007/978-3-319-29854-2_8}
\BIBentrySTDinterwordspacing

\bibitem{hoque2018architecture}
S.~Hoque and A.~Miranskyy, ``Architecture for analysis of streaming data,'' in
  \emph{2018 IEEE International Conference on Cloud Engineering (IC2E)}.\hskip
  1em plus 0.5em minus 0.4em\relax IEEE, 2018, pp. 263--269.

\bibitem{CloudFun20:online}
\BIBentryALTinterwordspacing
{Cloud Functions - Overview: IBM}. [Online]. Available:
  \url{https://www.ibm.com/cloud/functions}
\BIBentrySTDinterwordspacing

\bibitem{fielding2000architectural}
R.~T. Fielding and R.~N. Taylor, \emph{Architectural styles and the design of
  network-based software architectures}.\hskip 1em plus 0.5em minus 0.4em\relax
  University of California, Irvine Doctoral dissertation, 2000.

\bibitem{AWSLambd42:online}
\BIBentryALTinterwordspacing
Aws lambda – serverless compute - amazon web services. [Online]. Available:
  \url{https://aws.amazon.com/lambda/}
\BIBentrySTDinterwordspacing

\bibitem{CloudFun61:online}
\BIBentryALTinterwordspacing
Cloud functions  |  google cloud. [Online]. Available:
  \url{https://cloud.google.com/functions}
\BIBentrySTDinterwordspacing

\bibitem{ApacheOp33:online}
\BIBentryALTinterwordspacing
{Apache OpenWhisk is a serverless, open source cloud platform}. [Online].
  Available: \url{https://openwhisk.apache.org/}
\BIBentrySTDinterwordspacing

\bibitem{OpenFaaS:online}
\BIBentryALTinterwordspacing
Openfaas. [Online]. Available: \url{https://docs.openfaas.com/}
\BIBentrySTDinterwordspacing

\bibitem{CloudObj90:online}
\BIBentryALTinterwordspacing
Cloud object storage | store \& retrieve data anywhere | amazon simple storage
  service (s3). [Online]. Available: \url{https://aws.amazon.com/s3/}
\BIBentrySTDinterwordspacing

\bibitem{AzureBlo53:online}
\BIBentryALTinterwordspacing
Azure blob storage | microsoft azure. [Online]. Available:
  \url{https://azure.microsoft.com/en-us/services/storage/blobs/}
\BIBentrySTDinterwordspacing

\bibitem{ParseOpe49:online}
\BIBentryALTinterwordspacing
{Parse + Open Source}. [Online]. Available: \url{https://parseplatform.org/}
\BIBentrySTDinterwordspacing

\bibitem{OpenIOHi69:online}
\BIBentryALTinterwordspacing
Openio | high performance object storage for big data and ai. [Online].
  Available: \url{https://www.openio.io/}
\BIBentrySTDinterwordspacing

\bibitem{AmazonDy75:online}
\BIBentryALTinterwordspacing
Amazon dynamodb | nosql key-value database | amazon web services. [Online].
  Available: \url{https://aws.amazon.com/dynamodb/}
\BIBentrySTDinterwordspacing

\bibitem{Datastor73:online}
\BIBentryALTinterwordspacing
Firestore  |  google cloud. [Online]. Available:
  \url{https://cloud.google.com/firestore}
\BIBentrySTDinterwordspacing

\bibitem{ApacheCo24:online}
\BIBentryALTinterwordspacing
Apache couchdb. [Online]. Available: \url{https://couchdb.apache.org/}
\BIBentrySTDinterwordspacing

\bibitem{Themostp23:online}
\BIBentryALTinterwordspacing
The most popular database for modern apps | mongodb. [Online]. Available:
  \url{https://www.mongodb.com/}
\BIBentrySTDinterwordspacing

\bibitem{EventStr87:online}
\BIBentryALTinterwordspacing
{IBM Event Streams - Overview: IBM}. [Online]. Available:
  \url{https://www.ibm.com/cloud/event-streams}
\BIBentrySTDinterwordspacing

\bibitem{ApacheKa3:online}
\BIBentryALTinterwordspacing
{Apache Kafka}. [Online]. Available: \url{https://kafka.apache.org/}
\BIBentrySTDinterwordspacing

\bibitem{IBMCloud91:online}
\BIBentryALTinterwordspacing
{IBM Cloud: Kubernetes}. [Online]. Available:
  \url{https://cloud.ibm.com/kubernetes/overview}
\BIBentrySTDinterwordspacing

\bibitem{IBMAPIEx40:online}
\BIBentryALTinterwordspacing
{IBM API Explorer : Cloud Functions}. [Online]. Available:
  \url{https://developer.ibm.com/api/view/cloudfunctions-prod:cloud-functions:title-Cloud_Functions}
\BIBentrySTDinterwordspacing

\bibitem{IBMAPIEx60:online}
\BIBentryALTinterwordspacing
{IBM API Explorer Catalog}. [Online]. Available:
  \url{https://developer.ibm.com/api/list}
\BIBentrySTDinterwordspacing

\bibitem{AboutIBM67:online}
\BIBentryALTinterwordspacing
{About IBM Cloud Code Engine}. [Online]. Available:
  \url{https://cloud.ibm.com/docs/codeengine?topic=codeengine-about}
\BIBentrySTDinterwordspacing

\bibitem{IBMCloud29:online}
\BIBentryALTinterwordspacing
{IBM Cloudant - Overview}. [Online]. Available:
  \url{https://www.ibm.com/ca-en/marketplace/database-management}
\BIBentrySTDinterwordspacing

\bibitem{Limits67:online}
\BIBentryALTinterwordspacing
{IBM Cloudant: Limits}. [Online]. Available:
  \url{https://cloud.ibm.com/docs/Cloudant?topic=Cloudant-limits}
\BIBentrySTDinterwordspacing

\bibitem{CloudObj83:online}
\BIBentryALTinterwordspacing
{Cloud Object Storage - Pricing: IBM}. [Online]. Available:
  \url{https://www.ibm.com/cloud/object-storage/pricing/}
\BIBentrySTDinterwordspacing

\bibitem{stolwijk1999studying}
A.~Stolwijk, H.~Straatman, and G.~Zielhuis, ``Studying seasonality by using
  sine and cosine functions in regression analysis.'' \emph{Journal of
  Epidemiology \& Community Health}, vol.~53, no.~4, pp. 235--238, 1999.

\bibitem{hawkins1980identification}
\BIBentryALTinterwordspacing
D.~M. Hawkins, \emph{Identification of Outliers}, ser. Monographs on Applied
  Probability and Statistics.\hskip 1em plus 0.5em minus 0.4em\relax Springer,
  1980. [Online]. Available: \url{https://doi.org/10.1007/978-94-015-3994-4}
\BIBentrySTDinterwordspacing

\bibitem{hodge2004survey}
V.~Hodge and J.~Austin, ``A survey of outlier detection methodologies,''
  \emph{Artificial intelligence review}, vol.~22, no.~2, pp. 85--126, 2004.

\bibitem{zhang2017time}
A.~Zhang, S.~Song, J.~Wang, and P.~S. Yu, ``Time series data cleaning: From
  anomaly detection to anomaly repairing,'' \emph{Proceedings of the VLDB
  Endowment}, vol.~10, no.~10, pp. 1046--1057, 2017.

\bibitem{hochenbaum2017automatic}
J.~Hochenbaum, O.~S. Vallis, and A.~Kejariwal, ``Automatic anomaly detection in
  the cloud via statistical learning,'' \emph{arXiv preprint arXiv:1704.07706},
  2017.

\bibitem{wang2011solar}
Z.~Wang, F.~Wang, and S.~Su, ``Solar irradiance short-term prediction model
  based on bp neural network,'' \emph{Energy Procedia}, vol.~12, pp. 488--494,
  2011.

\bibitem{lipton2015critical}
Z.~C. Lipton, J.~Berkowitz, and C.~Elkan, ``A critical review of recurrent
  neural networks for sequence learning,'' \emph{arXiv preprint
  arXiv:1506.00019}, 2015.

\bibitem{hochreiter1997long}
S.~Hochreiter and J.~Schmidhuber, ``Long short-term memory,'' \emph{Neural
  computation}, vol.~9, no.~8, pp. 1735--1780, 1997.

\bibitem{cho2014learning}
K.~Cho, B.~Van~Merri{\"e}nboer, C.~Gulcehre, D.~Bahdanau, F.~Bougares,
  H.~Schwenk, and Y.~Bengio, ``Learning phrase representations using rnn
  encoder-decoder for statistical machine translation,'' \emph{arXiv preprint
  arXiv:1406.1078}, 2014.

\bibitem{islam2020anomaly}
\BIBentryALTinterwordspacing
M.~S. Islam and A.~V. Miranskyy, ``Anomaly detection in cloud components,'' in
  \emph{13th {IEEE} International Conference on Cloud Computing,
  {CLOUD}}.\hskip 1em plus 0.5em minus 0.4em\relax {IEEE}, 2020, pp. 1--3.
  [Online]. Available: \url{https://doi.org/10.1109/CLOUD49709.2020.00008}
\BIBentrySTDinterwordspacing

\bibitem{mohammads_thesis}
M.~S. Islam, ``{Anomaly Detection in Cloud Components},'' Master's thesis,
  Ryerson University, Toronto, Ontario, Canada, 2020.

\bibitem{ahmad2017unsupervised}
S.~Ahmad, A.~Lavin, S.~Purdy, and Z.~Agha, ``Unsupervised real-time anomaly
  detection for streaming data,'' \emph{Neurocomputing}, vol. 262, pp.
  134--147, 2017.

\bibitem{hundman2018detecting}
K.~Hundman, V.~Constantinou, C.~Laporte, I.~Colwell, and T.~Soderstrom,
  ``Detecting spacecraft anomalies using lstms and nonparametric dynamic
  thresholding,'' in \emph{Proceedings of the 24th ACM SIGKDD International
  Conference on Knowledge Discovery \& Data Mining}, 2018, pp. 387--395.

\end{thebibliography}

\end{document}